\documentstyle[prl,aps,epsf]{revtex}

\begin{document}
\title{Temperature dependence of the interlayer
magnetoresistance of quasi-one-dimensional Fermi liquids
at the magic angles}
\draft
\author{Ross H. McKenzie\cite{email} and Perez Moses }
\address{School of Physics, University of New
South Wales, Sydney 2052, Australia}
\date{\today}
\maketitle
\widetext

\begin{abstract}
The interlayer magnetoresistance of a quasi-one-dimensional
Fermi liquid is considered for the case of a magnetic
field that is rotated within the plane
perpendicular to the most-conducting direction.
Within semi-classical transport theory dips in the
magnetoresistance occur at integer ``magic angles'' only 
when the electronic dispersion parallel to
chains is nonlinear.
If the field direction is fixed at one of the magic angles
and the temperature is varied then the resulting variation
of the scattering rate can lead to a non-monotonic
variation of the interlayer magnetoresistance with temperature.
Although the model considered here gives a good description
of some of the properties of the Bechgaard salts, 
(TMTSF)$_{2}$PF$_6$ for pressures less than 8 kbar and
(TMTSF)$_{2}$ClO$_4$
it gives a poor description of their properties when the
field is parallel to the layers and of
the intralayer transport.
\end{abstract}


\section{Introduction}

In spite of intensive research 
over the past decade
the nature of the metallic state in low-dimensional
strongly correlated materials is still poorly understood.
Widely studied materials include cuprate and organic
superconductors\cite{ish,wosnitza}.
Many of the properties of the cuprates cannot 
be understood within the Fermi liquid picture
that has so successfully described conventional metals \cite{liang}.
Although some properties of the  quasi-two-dimensional
molecular crystals, $\kappa$-(BEDT-TTF)$_2$X (Reference\cite{mck})
and the  quasi-one-dimensional Bechgaard salts\cite{jerome}
 (TMTSF)$_{2}$X
can be explained within a Fermi liquid framework others
cannot.
A particular challenge is understanding the
dependence of the magnetoresistance of
the  Bechgaard salts 
on the direction of the magnetic field,
especially 
(TMTSF)$_{2}$PF$_{6} $ under pressures of about 10 kbar \cite{dan0,chas}.
The different angular-dependent magnetoresistance effects
in quasi-one-dimensional metals 
are known as the Danner \cite{danner}, magic
angle (or Lebed) \cite{leb,boe,lebbak,nau,osa,kang},
and third angular effects\cite{third},
depending on whether the magnetic field is rotated
in the ${\bf a}-{\bf c}$, ${\bf b}-{\bf c}$, or ${\bf a}-{\bf b}$
plane, respectively.
(The most- and least-conducting directions
are the ${\bf a}$ and ${\bf c}$ axes, respectively).
The magic angle effect is the most poorly understood
of these effects and is the focus of this paper.
If $\theta$ is the angle between the magnetic field
and the ${\bf c}$ axis 
then
at the ``magic angles'' given by
\begin{equation}
\tan \theta ={b\over c} {p \over q}  \hspace{20mm} \pm p,q = 1,2,3,....
\label{magicangle0}
\end{equation}
where $ b $ and $ c $ are the lattice constants in the $ {\bf b} $ and
 ${\bf  c}$ directions, Lebed predicted dips
in the threshold field for formation of
a field-induced spin-density-wave \cite{leb}.
Although these dips are not observed \cite{boe},
features are seen, mostly at $p/q=1,2$ in
the torque \cite{nau} and in all components of
 the resistance \cite{chas,nau,osa,kang,dan,chas2,lee}.

A wide range of physical mechanisms have been proposed
to explain these 
effects including commensurability effects changing
the electron-electron scattering rate \cite{lebbak},
 semi-classical transport \cite{mak},
complicated band structures \cite{osa2,blundell},
 hot spots on the Fermi surface \cite{chaikin},
cold spots on the Fermi surface \cite{pm:unpub},
 electron-electron interactions \cite{yak1},
non-Fermi liquid effects \cite{sca},
and magnetic field induced changes in effective dimensionality \cite{yak}.

The properties of 
(TMTSF)$_{2}$PF$_{6} $ at 10 kbar 
are particularly difficult to understand.
For example, when the magnetic field is
perpendicular to the current direction the magnetoresistance
is smaller than when it is parallel,
the opposite of what one observes  in 
(TMTSF)$_{2}$ClO$_{4}$
and in conventional metals.
Recently, the temperature dependence of
the magnetoresistance when the field direction
was fixed at the first magic angle was measured \cite{chas}.
It was found to be non-monotonic: as the temperature
decreased down to $T_{min}$, the magnetoresistance
decreased, it increased until $T_{max}$, and then decreased.
It has recently been proposed that these two temperatures
actually represent phase transitions between metallic
and insulating phases \cite{yak}.  The magnetoresistance
of the quasi-two-dimensional metal
 $\alpha$-(BEDT-TTF)$_2$MHg(SCN)$_4$ [M = K,Rb,Tl] also
exhibits unusual temperature and angular dependence
\cite{mcken,qualls}.

The purpose of this paper is to clarify what
properties of the magic angle effects can only
be explained within a non-Fermi liquid framework
by seeing what effects can be explained within
Fermi liquid theory.
The interlayer magnetoresistance is calculated within
the framework of semi-classical transport theory.
It is found that if one takes into account the
finite bandwith along the most-conducting direction
then dips in the magnetoresistance are observed for
$p/q=1,2,3,...$\cite{mak}. Furthermore, if one assumes a simple
Fermi liquid form for the temperature dependence of
the scattering rate then at the magic angles
the interlayer magnetoresistance does have a non-monotonic
temperature dependence.  Hence, one should be cautious
about associating maxima and minima in the 
temperature dependence with metal-insulator 
transitions.  
However, the results obtained gives a poor description
of the observed properties when the field is close
to the ${\bf b}$-axis and of the resistivity within
the layers.

\section{Calculation of the interlayer conductivity}

\subsection{Semi-classical transport theory}

If the electronic dispersion relation is $\epsilon(\vec k)$
then the
electronic group velocity perpendicular to
the layers is
$v_z = {1 \over \hbar} {\partial \epsilon(\vec k)
\over \partial k_z}.$
The interlayer conductivity can be calculated 
by solving the Boltzmann equation in the relaxation
time approximation leading to Chambers' formula \cite{ashcroft}
\begin{equation}
\sigma_{zz} = {e^2 \tau \over 4 \pi^3}
\int { v_z(\vec k) \bar{v}_z(\vec k)}
\left(-{\partial f(\epsilon) \over
\partial \epsilon}\right) d^3\vec k \ ,
\label{eq:chambers}
\end{equation}
where $f(\epsilon)$ is the Fermi function and $\tau$
is the scattering time
which is assumed to be the same at all
points on the Fermi surface.
$\bar{v}_z(\vec k)$ is the electron
velocity averaged over
its trajectories on the Fermi surface,
\begin{equation}
\bar{v}_z(\vec k) = {1 \over \tau}\int_{-\infty}^{0}
\exp \left({t \over \tau}\right)
v_z(\vec k(t)) dt  
 \label{eq:vbar}
\end{equation}
where $\vec k(0) = \vec k$.
The time dependence of the wave
vector $\vec{k}(t)$ is found by
integrating the semi-classical
equation of motion
\begin{equation}
{d \vec k \over dt} =
-{e \over \hbar^2} \vec{\nabla}_k
\epsilon \times \vec B \ .
 \label{prop}
\end{equation}
If the temperature is sufficiently
low that $T << E_F$ then
${\partial f \over \partial \epsilon}$ in
Equation~(\ref{eq:chambers})
can be replaced by a delta function
at the Fermi energy and Equation~(\ref{eq:chambers})
becomes
\begin{equation}
\sigma_{zz} = {e^2 \tau \over 4 \pi^3}
\int v_z(\vec k) \bar{v}_z(\vec k)
\delta(E_F-\epsilon(\vec{k})) d^3\vec{k}\ .
\label{chambers2}
\end{equation}

\subsection{Dispersion relation along the chains}
\label{calcCond}

 In the tight binding approximation the dispersion relation
in an orthorhombic crystal
can be written as
\begin{equation}
\epsilon (\vec{k})=-2t_{a}\cos (k_{x}a)-2t_{b}
\cos (k_{y}b)-2t_{c}\cos (k_{z}c) \ ,
\label{dispersion}
\end{equation}
where $ t_{a}, t_{b},$ and $ t_{a}$   are the
inter-site hopping integrals along the different crystal axes.
In the Bechgaard salts,
$ t_{a}\gg t_{b}, t_c $, 
the dispersion along the chains can be linearized
giving  
$ \epsilon (\overrightarrow{k})=\hbar v_{F}(|k_{x}|-k_{F})-2t_{b}
\cos (bk_{y})-2t_{c}\cos (bk_{z}) $,
where
$v_{F}=2t_{a}a\sin (ak_{F})/\hbar$ 
is the Fermi velocity and $k_F$ is the Fermi wave vector.
This linear dispersion has been used in a number
of papers on the magic angle effect \cite{osa2,blundell,chaikin,yak1}.
If one solves for the interlayer conductivity 
within semi-classical transport theory
one obtains

\begin{equation}
\frac{\sigma _{zz}(\theta )}{\sigma _{zz}^{0}}=
\frac{1}{1+(\omega_{c0}\tau \sin \theta)^{2}} ,
\label{nomagicangles}
\end{equation}
where $ \omega _{c0} = e v_F c B /\hbar $ is the frequency at which an electron
traverses the Brillouin zone in the ${\bf c} $-direction
when the field is parallel to the ${\bf  b}$-axis.
Clearly this is a smoothly varying function of $\theta$
and does not exhibit any magic angle effects.

We now show that if the full nonlinear dispersion
(\ref{dispersion}) is used then one does obtain 
magic angle effects. We will re-derive  a result obtained
earlier by Maki \cite{mak} with a view to   elucidating the physics
in the process.

\subsection{Solution of the semi-classical equations of motion}

The group velocity for the dispersion relation (\ref{dispersion})
is
\begin{equation}
\label{vel}
{\vec{v}(\vec{k})}=\frac{1}{\hbar }\overrightarrow{\nabla }_{k}\epsilon ={1\over
\hbar }
\pmatrix {2at_{a}\sin (ak_{x})\cr 2bt_{b}\sin (bk_{y})\cr
2ct_{c}\sin (ck_{z})} \ .
\end{equation}
The rate of change of the wave vector $ \vec{k}(t) $, in
a magnetic field in the ${\bf b}$-${\bf c}$ plane,
$ \vec{B}=(0,B\sin \theta ,B\cos \theta ) $, is
given by (\ref{prop}),
\begin{equation}
{d\vec{k}\over dt}=
{1\over \hbar ^{2}}\pmatrix {-2beBt_{b}
\cos \theta \sin (bk_{y})\cr {2aeBt_{a}\cos \theta \sin
(ak_{x})}\cr -2aeBt_{a}
\sin \theta \sin (ak_{x})}
%\hspace {100pt},
%\pmatrix {a\cr b\cr c} \  ,
\label{ktime}
\end{equation}
where terms involving $ t_{c} $ have been neglected.
This is valid provided $t_c \sin \theta \ll t_b \cos \theta$.
Hence, the results below will not be valid as $\theta \to 90^{0}$.

In order to calculate
the $ z $-component of the velocity one needs to obtain
$ \vec{k}_{z}(t) $, which Equation~(\ref{ktime}c) shows
is determined by $k_x(t)$.
To zero-th order in $t_b$, $k_x(t)= k_F$.
Integrating  Equation~(\ref{ktime}b) then gives
\begin{equation}
k_{y}(t)=k_{y}(0)+{\frac{\omega _{b}}{b}}t \  ,
\end{equation}
where
\begin{equation}
\omega _{b}=v_{F}eBb\cos \theta /\hbar \equiv \omega _{b0} \cos \theta
\end{equation}
is the frequency at which an the electron traverses the
Brillouin zone   in the direction of the ${\bf  b} $-axis.    

 Substituting this into Equation~(\ref{ktime}a) and
integrating
gives, to first order in $t_b/t_a$,
\begin{equation}
k_{x}(t)=k_{F}+{2t_{b}\over \hbar v_{F}}\cos (bk_{y}(0)+\omega _{b}t) \  .
\label{kxlebed}
\end{equation}
We obtain $ k_{z}(t) $ by using Equation~(\ref{ktime}c) and substituting in
(\ref{kxlebed}), giving

\begin{equation}
\label{wavevectlebed}
\frac{dk_{z}}{dt}=\frac{-2aeBt_{a}\sin \theta }{\hbar ^{2}}\left[ \sin
(k_{F})
\cos \left( \frac{2at_{b}}{\hbar v_{F}}\cos (bk_{y}(0)+\omega _{b}t)\right)
+
\cos (ak_{F})\sin \left( \frac{2at_{b}}{\hbar v_{F}}\cos (bk_{y}(0)+\omega
_{b}t)\right) \right] ,
\end{equation}
where we have used trigonometric identities to expand $ \sin (ak_{x}(t)) $.
If we take a linear dispersion relation, the second term in
(\ref{wavevectlebed})
will equal zero and we are left with $ \frac{dk_{z}}{dt}= $
$ \frac{-Bev_{F}\sin \theta }{\hbar } $
, where we have assumed that at $ t=0 $, the wave vector in the x-direction
($ k_{x} $) is equal to $ k_{F} $.

 Now to first   order in  $ {t_{b}}/{t_{a}}, $
\begin{equation}
\frac{dk_{z}}{dt}=\frac{-2aeBt_{a}\sin \theta }{\hbar ^{2}}
\left[ \sin (ak_{F})+\cos (ak_{F})\sin \left( \frac{2at_{b}}{\hbar v_{F}}
\cos (bk_{y}(0)+\omega _{b}t)\right) \right]  . \label{wavevectlebed2}
\end{equation}
Integrating this we obtain
\begin{equation}
k_{z}(t)c=k_{z}(0)c-\omega_{c}t-\gamma_0 \tan \theta
\sin (bk_{y}(0)+\omega_{b}t)\, \, ,
\label{kzt}
\end{equation}
where
\begin{equation}
\omega_{c}=\omega_{c0} \sin \theta 
\end{equation}
and
\begin{equation}
\gamma _{0}={2ct_{b}\over \hbar v_{F}}{a\over b}{\cot (ak_{F})} \ .
\label{ga1}
\end{equation}

\subsection{Evaluation of the interlayer conductivity         }

Substitution of (\ref{kzt}) into the $ z $-component of the velocity gives
\begin{equation}
v_{z}(k_{z}(0),\phi ,\phi ^{'})={2ct_{c}\over \hbar }
\sin \left( \hspace {4pt}ck_{z}(0)+{\omega _{c}
\over \omega _{b}}\phi ^{'}-\gamma _{0}\tan
\theta \sin (\phi -\phi ^{'})\right) \  ,
\label{velbarlebed}
\end{equation}
where
$ \phi ^{'}=-\omega _{b}t $, $ \phi =bk_{y}(0) $.

 The interlayer conductivity given by (\ref{chambers2})
can then be written in the form
\begin{equation}
\sigma _{zz}={e^{2}\over 4\pi ^{3}b\hbar v_{F}}
\int _{-\pi /c}^{\pi /c}dk_{z}(0)\int _{0}^{2\pi }d\phi \hspace {3pt}
{v_{z}(k_{z}(0),\phi )}\int _{0}^{\infty }{d\phi ^{'}\over \omega _{b}}
\exp \left( {\phi ^{'}\over \tau \omega _{b}}\right)
\hspace {3pt}v_{z}(k_{z}(0),\phi ,\phi ^{'}) \ .
\label{condlebed}
\end{equation}
We now expand Equation~(\ref{velbarlebed}) using trigonometric identies and
substitute
the Bessel generating functions to obtain
\begin{eqnarray}
v_{z}(k_{z}(0),\phi ,\phi ^{'}) & = & {2ct_{c}\over \hbar }
\left[ \sin \left( ck_{z}(0)+{\omega _{c}\over \omega _{b}}
\phi ^{'}\right) \left[ J_{0}(\gamma _{0}\tan \theta )+
2\sum _{k=1}^{\infty }J_{2k}(\gamma _{0}\tan \theta )\cos ((2k)
(\phi -\phi ^{'}))\right] \right. \nonumber \\
& & \left. +\cos \left( ck_{z}(0)+{\omega _{C}\over \omega _{b}}
\phi ^{'}\right) \left[ 2\sum _{k=0}^{\infty }J_{2k+1}(\gamma _{0}
\tan \theta )\sin ((2k+1)(\phi -\phi ^{'}))\right] \right]
\end{eqnarray}
and $ v_{z}(k_{z}(0),\phi ) $ is obtained by setting $\phi^{'}=0$.
Substituting
these expressions into Equation~(\ref{condlebed}) and performing the integrals
over $ \phi ^{'} $, $ \phi $ and $ dk_{z}(0) $ the final expression for
the conductivity becomes

\begin{equation}
\sigma _{zz}(\theta )=\sigma _{zz}^{0}\left[ {J_{0}
(\gamma _{0}\tan \theta )^{2}\over {1+(\omega _{c}\tau )^{2}}}
+\sum _{\nu =1}^{\infty }J_{\nu }(\gamma _{0}\tan \theta )^{2}
\left( {1\over {1+\tau ^{2}(\omega _{c}-\omega _{b}\nu )^{2}}}
+{1\over {1+\tau ^{2}(\omega _{c}+\omega _{b}\nu )^{2}}}\right) \right]  \
,
\label{condlebedfinal}
\end{equation}
where $ \sigma _{zz}^{0}={2e^{2}\tau ct_{c}^{2}\over \pi b\hbar ^{3}v_{F}} $
is the interlayer conductivity in zero field.
Note that for fixed $ \nu $ and $ z\ll 1 $
\begin{equation}
J_{\nu }(z)\approx {({z\over 2})^{\nu }\over \Gamma (\nu +1)}  .
\end{equation}
 Maki \cite{mak} obtained a similar
result, although he included the corrections to
$\omega_b$ and  $ \omega _{c} $
to next order
in $ \left( \frac{t_{b}}{t_{a}}\right) ^{2} $.
This raises the general question of to what order in
$t_b/t_a$ is the above expression for $\sigma_{zz}$ valid.
We only calculated $k_{z}(t)$ to
first order in $t_b/t_a$. Strictly speaking, this means that
(\ref{condlebedfinal})
is valid to second order in $t_b/t_a$.
However, we anticipate that a general solution for
$v_z(t)$ will be of the form
$v_z(t) \sim \sin (\omega_c t) \sum_n a_n \sin (\omega_b t)$
where $a_n $ is of order $(t_b/t_a)^n$.
This means that the coefficients in (\ref{condlebedfinal})
for $\nu \geq 2$ will change but be of the same order.

\section{Magic angles}

\begin{figure}
\centerline{\epsfxsize=12cm \epsfbox{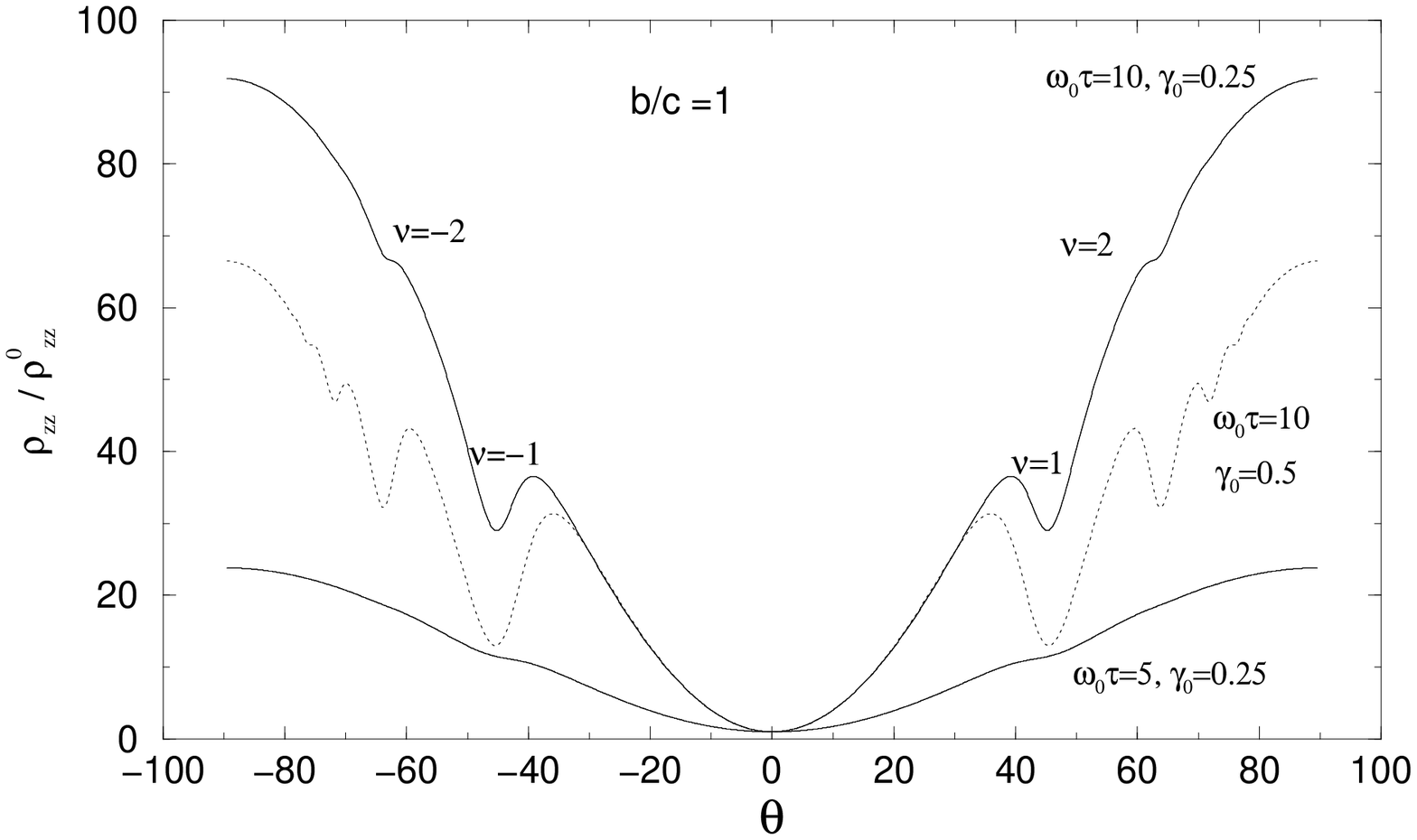}}
\caption{
Angular dependence of the interlayer magnetoresistance of a
quasi-one-dimensional
Fermi liquid. $ \theta$ is the angle between the magnetic field
and ${\bf c}$-axis and the magnetic field is rotated in the
 ${\bf b}-{\bf c}$
plane. The dips in the resistivity occur at the magic angles defined by
$ \tan \theta =\nu$
where, $ \nu =1 $ and $ 2 $ (see Equation~(\ref{magicangle})).
The dips only occur when one takes into account the
non-linear dispersion parallel to the chains
and their intensity depends on
 $ \gamma_{0} $,
which is determined by the geometry of the Fermi surface (Eq (\ref{ga1})).
 $ \tau  $ is the scattering time and $ \omega _{0} $
is the frequency at which the electron traverses the Fermi
surface when the field is perpendicular to the layers. 
The lattice constants $b$ and $c$ are set equal
and so $ \omega_{0} = \omega_{b0} = \omega_{c0}.$
The resisitivity is normalized to
$ \rho _{zz}^{0}$, the interlayer resistivity at zero
magnetic field.
\label{p4-1}} \end{figure}

The angular dependence of the interlayer
 resistivity given by Equation~(\ref{condlebedfinal}) is
 shown in Figure~\ref{p4-1} for several parameter values.
Dips occur at the ``magic angles'' given by $\omega_c = 
\nu \omega_b$ or
\begin{equation}
\tan \theta ={b\over c} \ \nu  \hspace{20mm} \pm \nu = 1,2,3,....
\label{magicangle}
\end{equation}
where $ b $ and $ c $ are lattice constants.
The size of the dip at the $\nu$-th magic angle,
compared the background magnetoresistance will be
of order
\begin{equation}
\left({\gamma_0 \over 2} {b\over c}  \nu \right)^{2 \nu} 
\left({ \omega_c \tau \over \nu! }\right)^2.
\label{size}
\end{equation}
The size of the dips is determined by the parameter $\gamma_0$,
defined by (\ref{ga1}), which is determined by
the geometry of the Fermi surface.
Note that if
 $ \gamma _{0}\rightarrow 0 $, (\ref{condlebedfinal}) reduces
to (\ref{nomagicangles}). This is because the limit $ \gamma _{0}
\rightarrow 0  $  
corresponds to taking a linear dispersion relation.
If $\gamma_0 \ll 1$ then the dips will decrease in
magnitude rapidly with increasing $\nu$.
For example if $\gamma_0 \sim  0.1$ 
then the $\nu = 1$ feature will be
five orders of magnitude smaller than the
$\nu = 3$ feature.
Note that when $\nu$ becomes sufficiently large this will
no longer be valid because $\gamma_0 \tan \theta \sim 1$.

We now consider what is a realistic value for $ \gamma _{0} $
for the  (TMTSF)$_{2}$X
materials. If we look at the
the form of
$ \gamma _{0} $ in Equation~(\ref{ga1}) we note that the factor
$ \frac{2ct_{b}}{\hbar v_{F}} $
equals the parameter $ \gamma $ which determines the periodicity 
of the Danner oscillations \cite{danner,mckenzie}.
For (TMTSF)$_{2}$ClO$_{4}$
it was estimated to be 0.24.
The lattice constants for  (TMTSF)$_{2}$PF$_{6} $ 
are $ a=7.3 \AA $, $ b=7.7 \AA $
and $ c=13.5 \AA $, while for
  (TMTSF)$_{2}$ClO$_{4}$,
$b$ is twice as large due to anion ordering \cite{wosnitza}.
 The $ \cot(ak_{F}) $
term depends on
the band filling. At three-quarter   filling
$ k_{F}=\frac{3\pi }{4a} $,
and $ \cot(ak_{F})=1 $. This gives a value
for $ \gamma _{0}(PF_{6})=0.24 $
and $ \gamma _{0}(ClO_{4})=0.12 $. Note that for half-filling 
$ k_{F}=\frac{\pi }{2a} $, and
thus $ \gamma _{0}=0 $ and there will be no magic angle effects
unless we solve the semi-classical equations to higher order
in $t_b$.

Figure~\ref{p4-1} is qualitatively similar to 
experimental results for  (TMTSF)$_{2}$ClO$_{4} $
 at ambient pressure \cite{dan}
and at 6.0 kbar \cite{chas2}
and (TMTSF)$_{2}$PF$_{6} $ at 6.0 kbar (0.3 K and 4 T) \cite{lee}.
A small difference is that the
experimental data shows a small dip near
$ 90^{0} $,
whereas the theoretical curve shows no such dip.
It is quite possible that the small dips can be explained within 
semi-classical transport theory if one includes the effect
of a finite $t_c$ in the solution of the semi-classical
transport equations.
An analogous effect occurs when the field is rotated in
the ${\bf a}-{\bf c}$ plane:
 for coherent interlayer transport with finite $t_c$
 a peak in the angular-dependent
magnetoresistance occurs when the field is
 parallel to the ${\bf a}$ axis \cite{hanasaki}.

The angular dependence of the interlayer magnetoresistance
of (TMTSF)$_{2}$PF$_{6} $
at pressures of about 10 kbar 
%and for fields larger than one tesla
is quite different
from that shown in Figure~\ref{p4-1}
At   fields less than one tesla the angular dependence
is similar to that given by Eqn. (\ref{nomagicangles}).
However, above one tesla,
$\rho_{zz} \sim (B \cos \theta)^{1.3}$
and so a large dip is observed near $ 90^{0} $
and at $ 90^{0} $ the in-field resistance is
comparable to the zero-field resistance \cite{chas}.

A number of theoretical papers \cite{lebbak,osa2,blundell,chaikin}
have predicted effects when 
$ \tan \theta ={b\over c} {p \over q} $
where $p/q$ is fraction. In contrast, the model
considered here only gives
effects for $q=1$. 
A review of the experimental literature shows 
that the only reproducible fractional features seen have been
in 
  (TMTSF)$_{2}$ClO$_{4}$
at $p/q=$ 3/2 and 5/2 \cite{nau}.          
This can be explained within the framework 
considered here.
If (TMTSF)$_{2}$ClO$_{4}$
is slowly cooled anion ordering occurs and the lattice
constant in the ${\bf b}$ direction doubles
so in (\ref{magicangle}) $b$ should be replaced by $2b$.
However, if a sample which is not completely anion ordered
it will produce features at angles corresponding to
half-integers for a fully anion-ordered sample.

\section{Temperature dependence of the interlayer magnetoresistance at the
magic angles}

Suppose that the field direction is fixed at a magic
angle and the temperature (and thus the scattering time $\tau$) is
varied.
%the conductivity can then be written as
%\begin{equation}
%\label{simpsig}
%\sigma _{zz}(\theta )=\sigma _{zz}^{0}\left[ {1\over {1+(\omega
%_{C}\tau )^{2}}}
%+\sum _{\nu =1}^{\infty }\left[ {({\gamma _{0}\tan \theta \over 2})^{\nu }
%\over \Gamma (\nu +1)}\right] ^{2}\left( {1\over {1+\tau ^{2}
%(\omega _{C}-\omega _{B}\nu )^{2}}}+{1\over {1+\tau ^{2}
%(\omega _{C}+\omega _{B}\nu )^{2}}}\right) \right]  \ .
%\end{equation}

\begin{figure}
\centerline{\epsfxsize=12cm \epsfbox{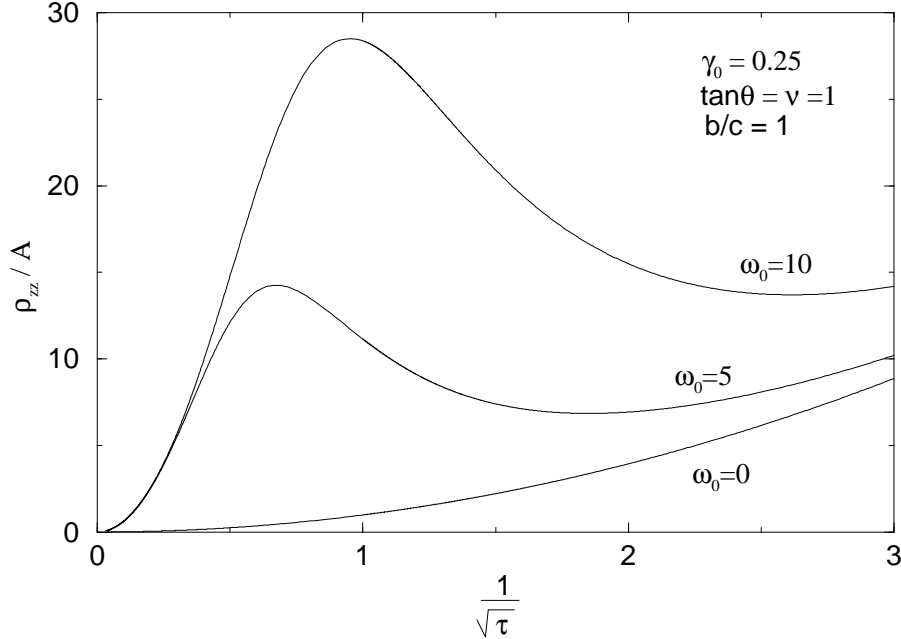}}
\caption{
Non-monotonic dependence of the interlayer resistivity,
at the $\nu =1$ magic angle,
on the scattering time $\tau$. 
The curves shown are for $\gamma_{0}=0.25$
and for various values of $\omega _{0}$, which is proportional
to the magnetic field. $1/\sqrt{\tau }$ is used for the
horizontal axis because it will an increasing function of
temperature in a Fermi liquid.
A local minima occurs at $\omega_{0}\tau \simeq {1\over \sin\theta_1}$
and there is a local maxima when $\omega_{0}\tau \simeq {2\over \gamma_0
\sin\theta_1}$.
The interlayer
resistivity $\rho _{zz}$ is normalised to ${1 \over A}$,
where $A$ is a constant equal to the ratio of the zero
field conductivity to the scattering time $\tau$.
\label{p4-2}}
\end{figure}

{\it The first magic angle} $ (\nu =1) $. 
Setting $ \omega _{c}=\omega _{b} $
and using the fact that
$ \gamma \ll 1 $ to    take   just the
first term in the series, i.e.
$ \nu =1 $, the conductivity   is
\begin{equation}
\sigma _{zz}(\theta _{1}) \simeq A\tau \left[ {1\over 1+
(\omega _{c0} \sin \theta_1 \tau )^{2}}+
\left( {\gamma _{0}\tan \theta _{1}\over 2}\right) ^{2}\right]  \  ,
\label{simpsig2}
\end{equation}
where $ \theta _{1} $ represents the first magic angle 
and $ \sigma _{zz}^{0}=A\tau $,
where $ A $ is the ratio of the zero field
conductivity to $ \tau $.
A plot of interlayer resistivity verses
$ {1\over \sqrt{\tau }} $ is shown
in Figure~\ref{p4-2} for different values
of $ \omega _{0} $. The interlayer
resistivity is a non-monotonic function
of $ \tau . $ It will be seen below that this leads
to non-monotonic temperature dependence.

 We now find for what
values of $ \tau $ the maxima and minima seen in Figure~\ref{p4-2}
occur. Finding the extrema of Equation~(\ref{simpsig2})
as a function of 
$ \tau $  gives that the minimun occurs when
\begin{equation}
\omega_{c0}\tau \simeq 
\frac{1}{\sin \theta_{1}} 
\label{min}
\end{equation}
and the maximum occurs when
\begin{equation}
\omega_{c0}\tau \simeq 
\frac{2}{\gamma_0 \sin \theta _{1}}.
\label{max}
\end{equation}
If $ \gamma _{0}\ll 1 $, then the maximum will only
be observed at sufficiently
high fields and in high purity samples.

%Contrast, the case of the minima for the Danner effect
%which occurs when the field is rotated in the a-c plane.
%.... I think the resistivity may be a monotonically increasing
%function of  $1/\tau$.

{\it The second magic angle} $(\nu =2) $. To obtain
 the interlayer conductivity
for the second magic angle 
then a  similar argument to that given above leads to
\begin{equation}
\label{magicang2sig}
\sigma _{zz}(\theta _{2}) \simeq
A \tau \left[ \frac{1}
{1+(\tau \omega _{c})^{2}}+\left( \frac{\gamma _{0}
\tan \theta _{2}}{2}\right) ^{2}\left(
 \frac{1} {1+\left( \frac{\tau \omega _{c}}{2}\right) ^{2}}
+  \frac{1} {1+\left( \frac{3\tau \omega _{c}}{2}\right) ^{2}}
\right) +\frac{\left( \gamma _{0}\tan \theta _{2}\right)
^{4}}{64}\right]  ,
\end{equation}
where we have set $ \omega _{c}=2\omega _{b} $ and $ \theta _{2} $
is the $ \nu =2 $ magic angle.
For small $\gamma_0$, the
minima is again given by $\omega_c \tau \simeq 1$.
To find the maxima we
 expand the first two terms in
(\ref{magicang2sig}) to fourth order in $ \frac{1}{(\tau \omega _{c})^{2}}.$
The maxima occurs when 
\begin{equation}
\omega_{c0}\tau \simeq 
\frac{4}{\sin \theta _{2}(\gamma _{0}\tan \theta _{2})^{2}}.
\label{max2}
\end{equation}
Since  this is smaller than (\ref{max}) by a factor
of $1/\gamma_0$, in order to see this maximum
even higher fields
and lower temperatures will be required than
for  the maximum associated with the
first magic angle.

{\it Conductivity as $ \theta \rightarrow 90^{0} $:} We
can expand the term
$ \left( \frac{1}{1+\tau ^{2}(\omega _{c}-\omega _{b}\nu )^{2}}+
\frac{1}{1+\tau ^{2}(\omega _{c}+\omega _{b}\nu )^{2}}\right) $
in the summation in (\ref{condlebedfinal}) to second order in $ \nu \cos
\theta $
to obtain

\begin{equation}
\frac{2}{1+a^{2}}(1+\nu ^{2}\cos \theta ^{2}A+......) \ ,
\end{equation}
where $ a=\omega _{0}\tau $ and $
A=\frac{a^{2}(3a^{2}-1)}{(1+a^{2})^{2}}\approx 3 $
for $ \omega _{0}\tau \gg 1 $. Substitution of this into the conductivity
gives
\begin{equation}
\frac{\sigma _{zz}(\theta \rightarrow 90^{0})}{\sigma _{zz}^{0}} \simeq
\frac{1}{(\omega _{0}\tau )^{2}}\left[ J_{0}(\gamma _{0}\tan \theta )^{2}
+2\sum _{\nu =1}^{\infty }J_{\nu }(\gamma _{0}\tan \theta )^{2}
(1+3\nu ^{2}\cos \theta ^{2}+...)\right]  \ .
\end{equation}
This can be simplified using the identities
\begin{equation}
\sum _{n=-\infty }^{\infty }J_{n}(z)^{2}=1\quad \quad \quad
\quad \sum _{n=-\infty }^{\infty }n^{2}J_{n}(z)^{2}=z^{2}/2 
\end{equation}
to give
\begin{equation}
\frac{\rho_{zz}(\theta = 90^{0})}{\rho_{zz}^{0}} \simeq
\frac{(\omega _{c0}\tau )^{2}}{1+3\gamma _{0}^{2}}.
\end{equation}
The resistivity is quadratic in field
 as for the case of a linear dispersion, but the co-efficient is smaller.
This is consistent with Figure~\ref{p4-1} which shows that the
resistivity near $90^{0}$ does decrease with increasing $\gamma_0$.
This field dependence is quite different to what is observed
in the  (TMTSF)$_{2}$X materials. 
For  (TMTSF)$_{2}$ClO$_{4} $ at 6.0 kbar \cite{chas2}
a linear field dependence is observed at high fields.
In (TMTSF)$_{2}$PF$_{6} $ at pressures from 6 to 10 kbar 
the resistivity saturates as the field increases \cite{chas,lee}.
However, caution is in order because derivation of
the quadratic dependence involved assuming that
$t_c \tan \theta \ll t_b$ and so we only expect a
quadratic dependence slightly away from $90^{0}$
or in the limit $t_c \to 0$.

{\it Fermi liquid model for the temperature dependence:} We now consider a
specific
model for the temperature dependence of
the scattering time $ \tau $. In a Fermi
liquid the scattering rate, at temperatures
much less than the Fermi temperature, has a temperature dependence of
the form \cite{yak,Gorkov} 
\begin{equation}
{1\over \tau }={1\over \tau _{0}}+\beta T^{2} \ ,
\label{fl}
\end{equation}
where the first term is due
 to impurity scattering and the second
is due to electron-electron scattering.
 Using this expression for $ \tau $ in
Equation~(\ref{simpsig2}) 
we  can now plot the temperature dependence
of the resistivity.                 This is shown in
Figure~\ref{p4-3}
for various values of $ \omega _{c0}\tau _{0} $.
The resistivity is not a monotonic function of
temperature but has a minimum when
$ \omega _{c} \tau (T) \sim 1$.
If the field is sufficiently high there is
also a maximum.
Using (\ref{fl}) we see that 
the minimum occurs at a temperature
\begin{equation}
 T_{min} \simeq \left({\frac{\omega _{c0}\sin \theta }{\beta }}\right)^{1/2} .
\label{tmin}
\end{equation}
If $ T_{max}^2 \gg 1/(\beta\tau_0)$ then
(\ref{max}) implies 
\begin{equation}
 T_{max} \simeq \left({\frac{\omega _{c0}\gamma _{0}
\sin \theta \tan \theta }{2\beta }}\right)^{1/2} .
\label{tmax}
\end{equation}

We are unaware of any measurements of the
temperature dependence of the interlayer
magnetoresistance of
(TMTSF)$_{2}$ClO$_{4}$ at the magic angles. 
Although the temperature dependence at the $\nu =1$
magic angle shown in Figure~\ref{p4-3} 
is similar to that reported in Ref. \onlinecite{chas}
for the {\it intralayer} resistance of
(TMTSF)$_{2}$PF$_6$ at 9 kbar,
the observed temperature dependence of the
interlayer
magnetoresistance is different \cite{chas3}.
It depends weakly on the temperature from
15 K down to about 3 K and then decreases.

The temperature dependence shown in Figure~\ref{p4-3} for
  $ \theta =90^{0} $ i.e., as the magnetic field is aligned with
the ${\bf b} $ axis the temperature dependence is
qualitatively similar to that observed in
(TMTSF)$_{2}$ClO$_{4}$ at ambient pressure \cite{ong}.
Again, qualitatively very different
behavior was 
observed \cite{chas} in (TMTSF)$_{2}$PF$_6$ at 10 kbar.
There it was found that the in-field resistance 
had a temperature dependence similar to
the zero-field resistance.

Zheleznyak and Yakovenko have given a heuristic argument
as to the origin of the maximum and minimum temperatures
seen in the intralayer resistance in Ref. \onlinecite{chas},
suggesting that metal-insulator and insulator-metal
phase transitions occur as the temperature
passes through these values \cite{yak}.
They argue that 
$ T_{min}\approx \omega _{b} \sim B$ and $ T_{max}\approx t_{c} $
which is independent of field.
Above $T_{min}$ the system is a two-dimensional metal.
Below  $T_{min}$, a
 magnetic field causes the electron motion in the ${\bf b}$
direction to be quantized resulting in a one-dimensional dispersion
and correlations producing insulating behavior.
Below $ T_{max}$, the interlayer coupling becomes important
and metallic behavior is recovered.
In contrast we find that $ T_{min} $ and
$ T_{max}$ are given by (\ref{tmin}) and (\ref{tmax}), respectively,
and both scale with $\sqrt{B}$.
Careful measurements should be able to distinguish between these
two different field dependences.

\section{Conclusions}

This paper only considers the interlayer resistivity $\rho_{zz}$,
whereas magic angle effects are also seen experimentally
 in the intralayer resistivity $\rho_{xx}$.
Maki pointed out that \cite{mak} the semi-classical theory will only give
resonances in $\rho_{xx}$ of order $(t_c/t_a)^2$ whereas they
are observed to be much larger.
A possible way around this problem is that experiments that
are meant to measure $\rho_{xx}$ may actually be measuring
some of $\rho_{zz}$. This is because in highly anisotropic
metals it is difficult to arrange the contacts and current
path so it lies completely within the layers.
This potential problem 
increases the motivation to make thin film samples of these
metals.

%what about torque?

It has been shown that
within semi-classical transport theory 
a nonlinear dispersion parallel to
the chains is necessary to produce dips in the
interlayer magnetoresistance at integer magic angles.
If the field direction is fixed at one of the magic angles
then one observes both minima and maxima in the
temperature dependence of the interlayer magnetoresistance.
This arises from the temperature dependence of the
scattering rate.  Since these maxima and minima can exist within
a Fermi liquid model one should be cautious about 
associating them with non-Fermi liquid behaviour or
metal-insulator transitions.  On the other hand, although
the Fermi liquid model considered here gives a good description
of many of the properties of 
 (TMTSF)$_{2}$PF$_6$ at pressures from 6 to 8 kbar
 and of (TMTSF)$_{2}$ClO$_4$
it gives a poor description of their properties when the
field is parallel to the layers and of
the intralayer transport.
Qualitatively very different behavior
is observed in
 (TMTSF)$_{2}$PF$_6$ at pressures of about 10 kbar;
explaining it remains a considerable theoretical challenge.

\acknowledgements

This work was supported by the Australian Research Council.
We thank P.M. Chaikin and M.J. Naughton for very helpful discussions.
We also thank P.M. Chaikin for sending us unpublished experimental data.

\vspace{0.3cm}

\begin{figure}
\centerline{\epsfxsize=12cm \epsfbox{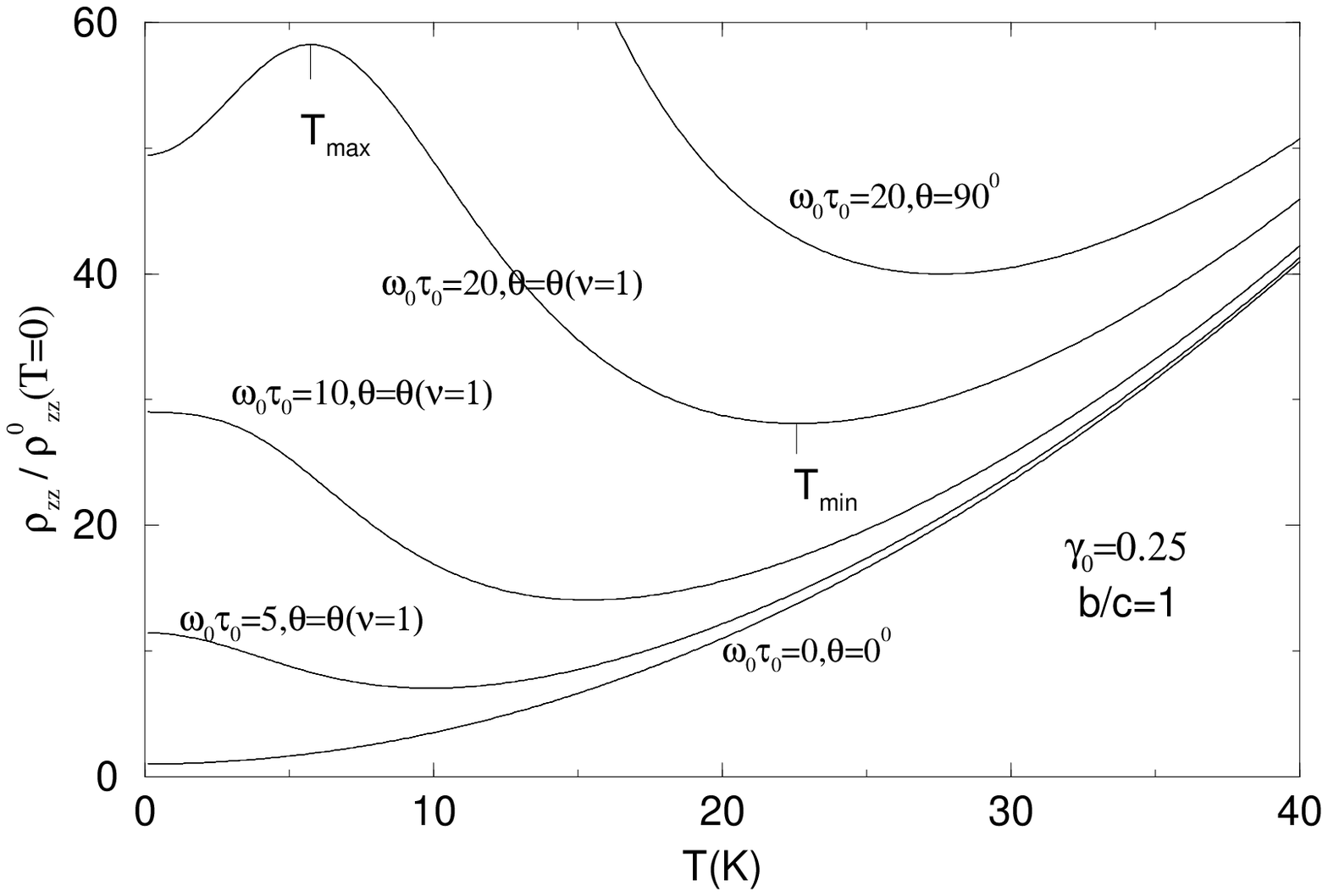}}
\caption{
Non-monotonic dependence of the
interlayer resistivity, at the $ \nu =1 $
magic angle, on temperature. A Fermi liquid form for the temperature
dependence
of the scattering rate is assumed.
A value of $ \tau_{0}\beta =0.025 $ K$^{-2}$ 
is used in Equation~(\ref{fl})
 so that the temperature
dependence of the zero-field resistivity roughly corresponds
to that of typical samples of Bechgaard salts.
At a temperature $ T_{min} $ defined by
 $ \omega _{0}\tau (T_{min})\simeq 1 $
there is a minimum in the resistivity. For sufficiently high fields there is
a temperatute $ T_{max} $ at which the resistivity is a maximum.
\label{p4-3}
}
\end{figure}


\begin{references}

\bibitem[*]{email}New address: Department
of Physics, University of Queensland, St. Lucia, 
4072, Australia; email: mckenzie@physics.uq.edu.au

\bibitem{ish} T. Ishiguro, K. Yamaji,
and G. Saito,
{\it Organic Superconductors}, Second Edition
(Springer, Berlin, 1998).

\bibitem{wosnitza}
J. Wosnitza,
{\it Fermi Surfaces of Low Dimensional
Organic Metals and Superconductors}
(Springer, Berlin, 1996).

\bibitem{liang}
For a recent experimental review, see 
W. Y. Liang,
J. Phys.: Condens. Matter  {\bf 10}, 11365 (1998).

\bibitem{mck}
R. H. McKenzie, Comments Cond. Mat. Phys. {\bf  18}, 309 (1998).

\bibitem{jerome}
C. Bourbonnais and  D. Jerome,
cond-mat/9903101.

\bibitem{dan0} G. M. Danner and P. M. Chaikin,
Phys. Rev. Lett. {\bf 75}, 4690 (1995).

\bibitem{chas}E. I. Chashechkina and P. M. Chaikin, Phys. Rev. Lett.
{\bf{80}}, 2181 (1998). 

%%%%%%%%%%%%%%%%%%%%%%%%%%%%%%%%%%%%%%%
\bibitem{danner} G. M. Danner, W. Kang, and P. M. Chaikin,
Phys. Rev. Lett. {\bf 72}, 3714 (1994).

\bibitem{leb}A. G. Lebed, Pis'ma Zh. Eksp. Teor. Fiz. {\bf{43}}, 137 (1986)
[JETP Lett. {\bf 43} 174 (1986)].

\bibitem{boe}G. S. Boebinger, G. Montambaux, M. L. Kaplan, R. C. Haddon, S.
V. Chichester
and L. Y. Chiang, Phys. Rev. Lett. {\bf{64}}, 591 (1990).

\bibitem{lebbak}A. G. Lebed and P. Bak, Phys. Rev. Lett. {\bf{63}}, 1315
(1989).

\bibitem{nau}M. J. Naughton. O. H. Chung and M. Chaparala, Phys. Rev. Lett.
{\bf{67}}, 3712 (1991).

\bibitem{osa}T. Osada, A. Kawasumi, S. Kagoshima, N. Miura and G. Saito,
Phys. Rev. Lett.
{\bf{66}}, 1525 (1991).

\bibitem{kang}
W. Kang, S. T. Hannahs, and  P. M. Chaikin,
 Phys. Rev. Lett. {\bf{69}}, 2827 (1992).

\bibitem{third}
T. Osada, S. Kagoshima, and N. Miura,
Phys. Rev. Lett. {\bf 77}, 5261 (1996);
A. G. Lebed and N. N. Bagmet,
Phys. Rev. B {\bf 55}, 8654 (1997);
I. J. Lee and M. J. Naughton,
ibid. {\bf 57}, 7423 (1998).


%\bibitem{mur}K. Murata, H. Bando, K. Kajimura, T. Ishiguro, H. Anzai, S.
%Kagoshima and G.
%Saito, Mol. Cryst. Liq. Cryst. {\bf{119}}, 131 (1985).

\bibitem{dan}G. Danner, W. Kang, 
and P. M. Chaikin, Physica B {\bf{201}}, 442 (1994).

\bibitem{chas2}E. I. Chashechkina and P. M. Chaikin, Phys. Rev. B.
{\bf{56}}, 13658 (1997).

\bibitem{lee}I. J. Lee and M. J. Naughton, Phys. Rev. B. {\bf{58}}, R13343
(1998).

\bibitem{mak}K. Maki, Phys. Rev. B. {\bf{45}}, 5111 (1992).

\bibitem{osa2}T. Osada, S. Kagoshima and N. Miura, Phys. Rev. B. {\bf{46}},
1812 (1992).

\bibitem{blundell}
S. J. Blundell and J. Singleton,
Phys. Rev. B {\bf 53}, 5609 (1996).

\bibitem{chaikin}
P. M. Chaikin, Phys. Rev. Lett. {\bf{69}}, 2831 (1992).

\bibitem{pm:unpub}
P. Moses and R. H. McKenzie, unpublished.

\bibitem{yak1}V. M. Yakovenko, Phys. Rev. Lett. {\bf{68}}, 3607 (1992).


\bibitem{sca}
S. P. Strong, D. G. Clarke, and P. W. Anderson,
Phys. Rev. Lett. {\bf 73}, 1007 (1994);
D. G. Clarke and S. P. Strong, Adv. Phys.
{\bf 46}, 545 (1997);
D. G. Clarke, S. P. Strong, P. M. Chaikin and E. I. Chashechkina,
Science {\bf 279}, 2071 (1998).

\bibitem{yak}A. T. Zheleznyak and V. M. Yakovenko,
Eur. Phys. J. B {\bf 11}, 385 (1999).

\bibitem{mcken} R. H. McKenzie, J. S. Qualls, S. Y. Han
and J. S. Brooks, Phys. Rev. B {\bf 57}, 11854 (1998).

\bibitem{qualls} J. S. Qualls, J. S. Brooks, S. Uji,
T. Terashima, C. Terakura, H. Aoki and L. K. Montgomery,
cond-mat/0005202.

\bibitem{ashcroft}N. W. Ashcroft and N. D. Mermin,
{\it Solid State Physics} (Saunders, Philadelphia, 1975).

\bibitem{mckenzie}
R. H. McKenzie and P. Moses,
Phys. Rev. Lett. {\bf 81}, 4492 (1998).

\bibitem{hanasaki}
N. Hanasaki, S. Kagoshima, T. Hasegawa, T. Osada, N. Miura,
Phys. Rev. B {\bf 57}, 1336 (1998).

\bibitem{Gorkov} L. P. Gorkov,
J. Phys. I. France {\bf 6},   1697, (1996);
L. P. Gorkov and M. Mochena,
Phys. Rev. B. {\bf{57}}, 6206  
(1998).
In a quasi-two-dimensional system there are
logarithmic corrections to (\ref{fl}).
However, such corrections have no qualitative effect on the
results presented here.

\bibitem{chas3}E. I. Chashechkina and P. M. Chaikin,
Synth. Metals {\bf 103}, 2176, (1999).

\bibitem{ong} G. M. Danner, N. P. Ong, and P. M. Chaikin,
Phys. Rev. Lett. {\bf 78}, 983  (1997).
At low temperatures $\rho_{zz}$ reaches a maximum value,
contrary to Figure~\ref{p4-3}. However, this maximum can be interpreted
as due to the presence of a high field superconducting
state [M. J. Naughton, I. J. Lee,
G. M. Danner, and P. M. Chaikin,
Synth. Metals {\bf 85}, 1481 (1997)].


\end{references}
\end{document}